\documentclass[preprint,prd,12pt]{revtex4}%
\usepackage{amsfonts}
\usepackage{amsmath}
\usepackage{amssymb}
\usepackage{graphicx}%
\providecommand{\U}[1]{\protect\rule{.1in}{.1in}}

\begin{document}
\preprint{quant-ph}
\title[Short title for running header]{Re-Creation: A possible interpretation of quantum indeterminism}
\author{M.B. Altaie}
\affiliation{Department of Physics, Yarmouk University, 21163 Irbid, Jordan}
\pacs{PACS number}

\begin{abstract}
I discuss some of the main interpretations given to explain the
indeterministic nature of quantum measurements and show that all has some
loopholes in one corner or another. I propose an alternative interpretation
based on the notion of continued re-creation of the physical properties. The
rate at which the system is re-created is a measure of the determinism of the
measurements. The existence of uncertainties is better explained through this
view, the meaning of incompatible observables becomes clearer, and with the
notion of re-creation and the origin of the Heisenberg uncertainty principle
becomes more vivid.

\end{abstract}

\volumenumber{number}
\issuenumber{number}
\eid{identifier}
\date[Date text]{date}
\startpage{1}
\endpage{102}
\maketitle
\tableofcontents

\section{Introduction}

It is a common understanding among physicists that the concept of quantum
measurement is still a problem which is in need for a solution in order to
clarify the deep implications of quantum theory. There is no consensus among
physicists; instead we have different views and interpretations on how quantum
measurements can be interpreted. Quantum measurements are the backbone of
applied quantum mechanics and, therefore it will be necessary to resolve this
problem for further development of quantum theory. Since the early years of
quantum mechanics this problem was the subject of fierce discussions and
controversy, all connected to one basic question and that is: how do we
interpret quantum mechanics?  \newline In this paper I will present a new
interpretation of quantum measurement based on the notion of continued
re-creation of the physical properties of systems. This notion may have its
early realization in opinions of some Greek philosophers, but surely had its
most sophisticated conceptualization in Islamic thought through the works of
Mutakallimun \cite{Wolfson}. Using this notion I will try to explain some of
the basic principles of quantum mechanics, at least on the conceptual level at
this stage, avoiding much mathematical details. Then I will try to foresee
some of the physical implications of such an interpretation and will glimpse
upon the philosophical and theological implications.

\section{Early development of quantum theory}

The discovery of the wave properties of particles, the particle properties of
waves, and the discreteness of many Observables in the atomic realm has
established the need for a new description of entities of the microscopic
world. At the beginning of the twentieth century many basic problems in atomic
physics were addressed leading to the establishment of quantum mechanics as a
paradigm to explain the observed properties of the atomic realm. The most
fundamental notions of early quantum mechanics were based on the assumption
that particles behave like waves. The main difficulty in realizing a wave-like
description for the particles lays in the fact that particles are localized
whereas waves are extended. This problem was overcome by the Louis de Broglie
suggestion that a particle can be represented by a wave which has a wavelength
inversely proportional to its momentum. This notion was soon utilized to
obtain a description of particles in terms of a de Broglie wave-packet with
the wavelength being that of a group of waves representing the particle. This
description opened the way to formulate the classical localized particle
mechanics in terms of the wave mechanics. Accordingly, Erwin Schr\"{o}dinger
formulated a wave equation in 1926 to describe the time development of atomic
particles under field of force \cite{Schrodinger}. The need to consider high
speeds required introducing the special relativistic formulation of the
problem and led to the well-known Dirac equation of the electron which was
discovered few years later \cite{Dirac}.

In essence the wave-like description of atomic particles benefited from all
the properties of wave phenomena and it was soon realized that the microscopic
world enjoys some basic properties that makes it different from the
macroscopic world. Particles, like atoms and electrons, are now being
identified as \textquotedblleft quantum states\textquotedblright\ symbolized
by the wave function $\psi(x,y,z,t)$. This is a mathematical expression
summarizing the physical content of a physical system in terms of spacetime
coordinates and other parameters of the system like energy and momentum. The
mathematical nature of $\psi(x,y,z,t)$ was already recognized since the early
days of formulating the Schr\"{o}dinger equation, and it was realized that the
wave function has no direct physical meaning in itself. Soon Max Born
\cite{Born1} was able to identify%

\begin{equation}
|\psi^{\ast}(x)\psi(x)|=|\psi(x)|^{2},
\end{equation}
to stand for the probability density of finding the particle in the position
$x$.

The wave-mechanical description of particles set by Schr\"{o}dinger was best
realized by saying that a particle is a wave-packet that is composed by
super-posing many basic (plain) waves. This description soon faced many
difficulties. The slightest dispersion in the medium will pull the wave-packet
apart in the direction of propagation, and even without such dispersion it
will always spread more and more in the transverse direction. Because of this
blurring a wave-packet does not seem to be very suitable to represent a
particle. Shortly before Schr\"{o}dinger had formulated his wave equation,
during the early summer of 1925, Werner Heisenberg \cite{Heisenberg} conceived
the idea of representing physical quantities by sets of complex numbers. This
was soon elaborated by Born, Jordan and Heisenberg \cite{Born2} himself into
what has become known as \emph{matrix mechanics}, the earliest consistent
theory of quantum phenomena. Both views, the wave mechanics of Schr\"{o}dinger
and the matrix mechanics of Heisenberg are said to be equivalent despite
differences in some basic concepts and formulation.

Few years later Jon von Neumann \cite{Neumann} formulated as a calculus of
Hermitian operators in Hilbert space. The wavefunction was represented by
complex function in an infinite-dimensional space covered by basis vectors.
According to the formalism set by von Neumann a physical system is completely
described by a wave function $|\psi>$, which is now to be taken as a vector in
an infinite-dimensional Hilbert space. A measurement of any observable $a$
belonging to the system is the result of the action of a mathematical operator
$\hat{A},$ corresponding to that observable, on the state vector (wave
function) representing the system. The result of such an operation is to
produce a value (a number called the eigenvalue) that stands for the value of
the observable at the moment of measurement. With this new comprehension
natural objects, which objectively were identified as ontologically existing
things, became known in terms of new epistemological entities that are
represented by abstract mathematical forms. It should be emphasized that this
is a very important turning-point in the history of scientific thought. The
fact that $|\psi>$ which represent the physical system is a mathematical
expression that has no direct physical meaning as noted earlier and the fact
that physical observables became obtainable in the theory only as a result of
operating certain mathematical operators on $|\psi>$ is surely a clear
indication of the fundamental turn that was implied by quantum mechanics.

The quantity $a$ (the observable) cannot be taken as such to stand for the
physical value of the observable; it has to averaged within the state of the
system and it is then called the \emph{expectation value} of the operator
$\hat{A}$ at the state $|\psi>$. This is the average value of all possible
measurements that can be carried in the system in the state $|\psi>$. However
we have to remember that theoretically the number of all possible measurements
is infinite, for this reason the expectation value may not be obtained in any
single measurement. 

\section{The Heisenberg uncertainty principle}

Much to the curiosity of the physicists, some aspects of the wave-like
description of particles led to some uncertainties in determining
simultaneously pairs of observables like position and momentum, energy and
time and other observables. This was expressed by the Heisenberg uncertainty
principle which, in one of its forms state that the position of a particle and
its momentum can never be determined simultaneously with infinite accuracy.
This principle contributed to the indeterminacy of the quantum world and had
taken much attention and interest from physicist. The uncertainty principle is
deeply rooted in the wave-mechanical description of particles; once we
represent a particle by a wave then it is inevitable that we should allow for
some kind of a distribution of the values of its position and momentum. The
Fourier analysis of such a description shows that the wave description
requires some inevitable non-locality in position which leads to the inherent
uncertainty in these variables. Similar situation applies to measurement of
time where it would lead to mutual uncertainty between time intervals and the
corresponding energies. The uncertainty relations are related to the
non-commutativity of the respective operators. In matrix mechanics operators
are matrices and in such case%

\begin{equation}
a\times b\neq b\times a,
\end{equation}
for this reason the operators of position and momentum do not commute.
Likewise are the operator for time and the Hamiltonian, which is the operator
for energy. This in turn will eliminate the possibility of finding a
simultaneous eigenvectors for the position and momentum. Instead we relate the
two separate eigenvectors by a Fourier transform. It is important to note that
indeterminacy of position and momentum caused tremendous shock to classical
physicists. The classical equation of motion of a particle requires knowing
both the initial position and the initial momentum. Having been denied such a
knowledge physicists were puzzled with the solutions of the equation of
motion. This caused the downfall of classical mechanics in the microscopic
world. The glory of classical mechanics, especially in its most sophisticated
form devised mainly by Lagrange and Hamilton still provoke some physicists to
re-establish the reign of classical physics.

\section{Discreteness and continuity}

The quantum indeterminacy problem is deeply rooted in the long-lasting
question of discreteness and continuity. This is an issue which has been under
persistent debate since the early days of the Greek, throughout the Islamic
period which witnessed fierce debates between the philosophers and
Mutakallimun \cite{Wolfson}.

The indeterminacy of quantum states described by the Heisenberg uncertainty
principle brought to the attention of physicists the fact that quantum
mechanics is a mechanics of the undetermined nature. As noted above, this soon
posed what came to be known as the \emph{measurement problem}\ in quantum
mechanics. This today, more than three quarter of a century after the advent
of the theory, is still an issue of unprecedented dissension. In fact it is by
far the most controversial problem of current research in the foundation of
physics and divides the community of physicists and philosophers of science
into numerous opposing schools of thought.

The main issues in this division seem to be centered on two things: the
quantum jumps, and the measurement indeterminacy. Quantum jumping is an
indication of the discrete nature of the atomic world. If this is to be a
fundamental characteristic of the microscopic world then continuity of the
macroscopic world would seem to be only fictitious. It was reported that
Schr\"{o}dinger once said \textquotedblleft if all this damned quantum jumping
were really to stay I shall be sorry I ever got involved with quantum
theory\textquotedblright\ \cite{Jammer}. The main difficulty will arise when
we find that our differential calculus, which is the backbone of the
mathematical formulation of classical physics that was based on continuity and
infinite divisibility, will be in need of serious revision. Consequently, the
canonical formulations of physical laws will not be valid and the basic
concepts of field theory will be challenged. The Schr\"{o}dinger equation is a
deterministic equation that adopts the principle of continuity and the concept
of infinite divisibility. However, a wave equation has helped to provide an
approximated picture of the quantum world. The discrete features of the
quantum world are now being presented as product of the wave mechanical nature
which allows for superposition of waves producing interference pattern.
Consequently, one can avoid thinking of the abrupt quantum jumps in favor of
some more lenient thinking in terms of the probability distribution, such that
some kind of continuity between discrete states is maintained. Therefore,
instead of having the macroscopic continuity becoming an apparent feature that
hides the underlying discreteness, we now have discreteness appearing as an
emergent product of some phenomena of the continuum. Beside this, it would be
important to note that precise analysis of the quantum phenomena of the two
slit interference shows some fundamental characteristic departure from the
standard wave-interference phenomena \cite{Namiki}. In these experiments, a
particle remains to be non-divisible. However, such a departure awaits an
explanation, which can precisely identify those features in both phenomena
that makes them different.

\section{The applicability quantum mechanics}

In this context comes the question whether quantum mechanics is a theory that
can be applied to a single particle or is it a theory of ensembles. Physicists
have different opinions on this issue. Some of them, like Bohr and Heisenberg,
believe that quantum mechanics is suitable to describe single particles as
well as many particle systems. This is generally the view held by the
Copenhagen school. Others physicists, like Einstein and Born, believe that
quantum mechanics is only applicable to ensembles rather than individual
particles, and accordingly it can only be interpreted statistically. Others,
like Everett and Wheeler, believe that quantum mechanics is essentially an
interaction theory that can be realized only through the interaction between
the observer and the system. In one way or another, this will allow for a
subjective interference in determining quantum states. In fact, the basic
formulation of the equation of motion in quantum mechanics, Schr\"{o}dinger's
equation, suggests that it can be applied to single particles, on the other
hand having the values of observables coming out as an average only may
suggest that we are talking about an ensemble of particles in which each
particle enjoys different value for that observable. The general behavior of
the system of these particles then is represented by the behavior of the
average. However, this restriction becomes unnecessary if we would interpret
the existence of an average as being happening as a result of many
measurements being performed on the same particle. In this case the implicit
fact will be that the value of the observable assigned to the system (the
single particle in this case) is not fixed but is ever changing. But then the
question arises as to whether this change in the value of the observable is
due to the changing state of the system, or is it due to the process of the
measurement itself. If we assume that it is due to the changing state of the
system then the process of measurement can be taken to be completely passive.
On the other hand, if it will be considered to be a result of the measurement
itself then we are assuming primarily that the measurement itself has a
disturbing effect on the system. This amounts to assume the existence of an
interaction between the system and the measuring device. Having the
microscopic systems being so small and delicate, no one can deny that such
possible interactions may cause subsequent disturbances. Therefore, such
interactions will lead to \emph{de-cohere} the quantum system. The
disturbances caused by the measuring devices are generally non-systematic and
so complicated that it would be unpredictable. On the other hand one might
expect that in some cases the disturbances caused by the macroscopic measuring
device can be so large that it will overwhelm the basic value of the
observable under measurement. The third point to make here is that such
disturbances, if known, can be accounted for in the equation of motion through
the potential term in the equation of motion. Accordingly, the case will
always be that of an interacting system for which the equation of motion may
be solved exactly or through numerical techniques. Virtually anything
environmental can be included in the potential of the system, which controls
the behavior of the system through the equation of motion. Considering these
notes, it would be odd to assume that quantum indeterminacy is a shear result
of the incision of measurement.

\section{Interpretations of quantum measurements}

In a given individual experiment, the result of the measurement is one of
several alternatives. A repetition of the experiment under identical initial
conditions may lead to another of these possible alternatives. This is
incompatible with the unitary evolution of Schr\"{o}dinger. Several solutions
have been proposed for this apparent inconsistency. The main ones are:

\subsection{The von Neumann Interpretation: wavefunction collapse}

To explain the process of measurement von Neumann suggested that the state
function changes according to two different ways (see for example
\cite{Neumann}): \newline\textbf{Process 1}: a discontinuous change brought
about by observations by which the quantity with eigenstate $|\psi>$ is
projected onto the state%

\begin{equation}
|\phi>=\hat{A}|\psi>
\end{equation}
instantly with probability%

\begin{equation}
\mid<\psi\mid\phi>\mid^{2},
\end{equation}
This amounts to determine the overlap between the state $|\psi>$ and the state%

\begin{equation}
|\phi>=\hat{A}|\psi>.\label{q1}%
\end{equation}
\newline\textbf{Process 2}: a change in the course of time development
according to the deterministic Schr\"{o}dinger equation. \newline The
description in process 1 is called \emph{the wavefunction collapse,} which
means that the state $|\psi>$ after measuring the observable $A$ will be
converted into the state $|\phi>$ given in Eqn. (\ref{q1}).

In this formulation of von Neumann a fundamental problem was recognize long
ago, this is the embodied apparent inconsistency between the indeterministic
nature of process 1 and the deterministic nature of process 2. This apparent
inconsistency has been presented in different forms, and it is in fact deeply
rooted in the formulation of quantum mechanics from its very beginning. Joseph
Jauch \cite{Jauch} presented the problem as follows: the problem of
measurement in quantum mechanics concerns the question whether the laws of
quantum mechanics are consistent with the acquisition of data concerning the
properties of quantum systems. This consistency problem arises because the
system to be measured and the apparatus which is used for the measurement are
themselves systems which are presumed to obey the laws of quantum mechanics.
Therefore the evolution of the state of such system is governed by the
Schr\"{o}dinger equation. However, the measuring process exhibits features,
which are apparently inconsistent with the Schr\"{o}dinger-type evolutions.
The typical process ends with the establishment of a permanent and
irreversible record. This contradicts the time-reversible Schr\"{o}dinger
equation. So, despite the fact that the von Neumann interpretation of quantum
measurement was adopted by the Copenhagen school, nevertheless it suffers from
some fundamental problems.

\subsection{The statistical interpretation}

For this we have two views \newline\textbf{Viewpoint I}: by which quantum
mechanics is understood to apply to ensembles and not to single particles.
Albert Einstein was an advocate of this interpretation. Einstein says: "The
function $\psi$ does not in any way describe a condition which could be that
of a single system: it relates rather to many systems, to `an ensemble of
systems' in the sense of statistical mechanics."\ \cite{Einstein}. Einstein
hoped that a future more complete theory may describe quantum mechanics as an
approximation of a more general one. \newline\textbf{Viewpoint II:} which was
proposed by Born, and supported by Bohr, according to which the wavefunction
$\psi$was understood to be symbolic of representation of the system and that%

\begin{equation}
|\psi(x)|^{2}=\psi^{\ast}(x)\psi(x),
\end{equation}
is taken to describe the probability density for the system is in the position
$x$. But probability can only be understood to have a meaning through a
population. In this case the population is that of many repeated measurements.
This may be asserted by the fact that Born was of the opinion that his
suggestion is of the same content as that of Einstein and that "the difference
[in their views] is not essential, but merely a matter of
language."\cite{Born3}. \newline One can say that the Einstein interpretation
is covered by the fact that in any measurement on a quantum system we measure
macroscopic quantities, a fact which was originally emphasized by Bohr. If,
however, we come to measure by any means a microscopic quantity then the
Einstein interpretation will not be valid. On the other hand, by requiring
that many measurements are to be done on the same system, Born's
interpretation implicitly assumes that the system is to remain within the same
state over the duration of all those measurements. Obviously this cannot be
generally guaranteed.

\subsection{The hidden variables interpretation}

This interpretation was championed by David Bohm \cite{Bohm} who assumed that
quantum mechanics is incomplete, and that there are some hidden variables that
should complement the physical description in order to get the full picture of
the physical world, which is assumed to be deterministic. There are several
kinds of hidden variable theories, some are local and some are non-local.
Belinfante \cite{Belinfante} has given a very detailed account of these
theories both in their scientific content and in their historical development.
By Bell's theorem \cite{Bell} the local hidden variable theories were shown to
be inconsistent with quantum mechanics. There remains to say that none of the
existing non-local theories is found to conclude any prediction that is new to
the standard formulation of quantum mechanics.

\subsection{The multi-world interpretation}

This was originally proposed by Hugh Everett \cite{Everett} in 1957. Everett
reformulated the process of measurement abandoning the concept of wavefunction
collapse set by process 1 of the von Neumann formalism, while keeping the
assumption of the deterministic evolution of the system under Schr\"{o}dinger
equation. Everett criticized the need for \textquotedblleft external
observers\textquotedblright\ to obtain measurements by the von Neumann scheme
and instead went to consider the system as being composed of two main
subsystems: the object and the measuring device (or observer). This
formulation established the concept of \textquotedblleft relative
state\textquotedblright. The treatment lead Everett to conclude that:
\textquotedblleft throughout all of a sequence of observation processes there
is only one physical system representing the observer, yet there is no single
unique state of the observer (which follows from the representations of
interacting systems). Nevertheless, there is a representation in terms of a
superposition, each element of which contains a definite observer state and a
corresponding system state. Thus, with each succeeding observation (or
interaction), the observer state {"}branches{"} into a number of different
states. Each branch represents a different outcome of the measurement and the
corresponding eigenstate for the object-system state. All branches exist
simultaneously in the superposition after any given sequence of observations".
Everett went further to suggest that: "the trajectory of the memory
configuration of an observer performing a sequence of measurements is thus not
a linear sequence of memory configurations, but a branching tree, with all
possible outcomes existing simultaneously in a final superposition with
various coefficients in the mathematical model. In any familiar memory device
the branching does not continue indefinitely, but must stop at a point limited
by the capacity of the memory". John Wheeler supported the Everett theory
emphasizing its self-consistency \cite{Wheeler}. An elaboration of the Everett
interpretation was also the subject of a study by Graham \cite{Graham} working
under the supervision of Bryce DeWitt. It was assumed that the eigenvalues
associated with the observer subsystem form a continuous spectrum, whereas the
eigenvalues associated with the object form discrete set. In order to
reconcile the assumption that the superposition never collapses with ordinary
experience which ascribes to the object system after the measurement only one
definite value of the observable, it was proposed that the world will be
splitting into many-worlds existing simultaneously where in each separate
world a measurement yield only one result, though this result differs in
general from one world to another.

\section{The re-creation postulate}

In order to interpret quantum measurement I propose the following two
postulates:\newline\textbf{Postulate P (1)}: All physical properties of a
system are subject to continued re-creation.\newline\textbf{Postulate P (2)}:
The frequency of re-creation is proportional to the total energy of the system.

It will be shown below that the re-created observables assumes a new value
every time it is re-created. This will cause the observable to have a
distribution of values over certain range (width) that is always controlled by
the re-creation frequency. The higher the total energy of the system the
narrower is the range of values over which the dispersion is expected and vise
versa. For this reason macroscopic systems are expected to behave classically,
whereas microscopic systems exhibit mostly quantum behavior. Clearly, the
narrower the dispersion of values, the more determinable is the value of the
observable and vice versa.

\section{Re-Creation and the Uncertainty Principle}

Once created an observable assumes a given basic value defined by the state of
the system at that moment. According to the re-creation postulate physical
parameters are permanently in a natural process of continued re-creation,
irrespective of the measurement operation. However, values of those parameters
can only be known at the time of measurement. Re-creation is a process of
change. Once a given parameter is re-created other parameters of the system
will be affected; thus changing their values in accordance with the related
physical laws. Any change is best described, in the most general form by the
generator corresponding to that parameter. For example if $x$ is re-created
then the system will change infinitesimally by $\partial/\partial x$ but this
is just proportional to the momentum operator. This will duly cause the value
of the position $x$ to change every time it is re-created, thus presenting a
distribution of values for $x$ instead of acquiring one single value.
Conversely if $p$ is re-created then the whole system will change by
$\partial/\partial p,$ but this will cause an infinitesimal shift in the value
of $p$ and consequently a shift in the value of the position parameter $x$.
Therefore every time an $x$ is re-created a change in the momentum of the
system will occur and conversely every time the momentum is re-created a
change in the value of the position will occur. This means that re-creating
the position will result in creating momentum and vice versa. If the system
itself is to stay invariant under the process of re-creation then we must have%

\begin{equation}
\left(  \frac{\partial}{\partial x}x-x\frac{\partial}{\partial x}\right)
|\psi>=|\psi>.
\end{equation}
Using the explicit forms for the position and momentum operators this would
imply that%

\begin{equation}
\hat{p}\hat{x}-\hat{x}\hat{p}=[\hat{p},\hat{x}]=-i\hslash,
\end{equation}

In other words, the effect of change is logically being seen as a commutation
of the parameter and its generator (which were also called complementary
observables). This is the well-known commutation relation that led to the
Heisenberg uncertainty relations. In this scheme however measurements could be
passive action that does not necessarily affect the system itself.

This proposal of re-creation preserves the statistical nature of the possible
values of the observables and resolves the question whether of quantum
mechanics is applicable to single particle or to an ensemble of particles.
Here we see that the single particle state is being under continued
re-creation, thus forming an ensemble of values on its own if a memory is to
be available to keep records of all values assumed under re-creation.
Nevertheless, a measurement of an observable taken over duration of time
exceeding the re-creation period will always yield an average of the values
assumed by the system during that period of measurement. So, practically we
almost measure average values every time we perform a measurement. This
explains how the probabilistic behavior arises in the case of single particle
quantum system. According to the above scheme we always measure average values
with very low dispersion for macroscopic objects; the re-creation frequency is
very high and consequently the measurement time cannot coup with the
re-creation period. This gives the macroscopic world its classical, apparently
deterministic, characteristics. This is why we have the measured values of the
observables of a macroscopic system always being very close, even identical,
to the theoretical expectation values of the observables. On the other hand in
microscopic systems the re-creation frequency is relatively low and,
therefore, we would expect the dispersion of values to be high enough exposing
the indeterministic character of the world.

The proposal also provides us with better understanding of the origin of the
uncertainty relations. Here we see that the appearance of uncertainty in the
values of complementary observables is a direct result of re-creation and the
entanglement of such variables. This means that indeterminism is a direct
consequence of the continued re-creation.

\section{Physical Implications of Re-creation}

There are several implications of the proposed re-creation scheme described
above. Some of these implications may be used to test the theory. However,
because of the mostly technical nature of these implications, I will only
provide an overview of those implications at this stage. The full technical
treatment of these implications might be presented elsewhere.

\subsection{Macroscopic quantum states}

The re-creation frequency can be affected by external field of force. Since it
is known from the theory of general relativity that any time duration for an
event occurring near a gravitational field of force is dilated by a factor
proportional to the strength of the field then one should expect that
re-creation periods are to be dilated once being in the vicinity of a strong
source of gravity, e.g., a compact astronomical object like white dwarfs and
neutron stars (see for example, \cite{Weinberg}). Consequently, re-creation
frequencies should be red-shifted once being in the vicinity of a strong
gravitational source. This means that macroscopic classical processes would
turn to exhibit quantum features once being in a strong gravitational field.
This will cause the appearance of macroscopic quantum states in such regions,
e.g. near the event horizon of black holes.

\subsection{Quantum coherence}

Coherence is one basic feature, which is realized in quantum systems, and it
is customary known that coherent systems are quantum systems. Such systems are
always featured with high efficiency e.g. lasers. The availability of
macroscopic quantum state may make it plausible to expect the occurrence of
macroscopic coherent states too, thus opining the way to understand some very
obscure phenomenon like the gamma-ray bursts which are known to occur at the
far rim of the universe. Beside this the re-creation postulate allows for a
new definition of coherence by which two systems can be considered coherent if
their re-creation frequencies are identical and their re-creation occurs in
the same phase.

\subsection{Quantum Zeno effect}

This is a very interesting proposal, which was suggested by Misra and
Sundarshan \cite{Sun} in 1977. The proposal is based on the notion of wave
function collapse and was considered to be a prediction of the collapse
interpretation. The idea is that if continuous measurements are carried on a
given state, then the system is expected to stay in that state because of the
continuous collapse of the wave function onto the same state. As they say, a
watched pot never boils. There was a claim that this prediction was verified
\cite{Itano}, but such claims were soon refuted \cite{Petrosky}. Recently some
more rigorous calculations have been done trying to present the quantum Zeno
effect (QZE) quantitatively in more accurate form taking into consideration
the effect of the measurement duration \cite{Schulman}. The re-creation
interpretation presented in this paper sets an upper limit for the measurement
time for the QZE to be possibly verified. The measurement time of observable
(say transition energy) should be less than the re-creation period for the QZE
to occur. Measurements performed within time durations, which are more than
the re-creation time will result in averaging the values of the observable
over several re-created states and consequently cannot hold the system at a
specific state, consequently QZE will not be verifiable in such cases.

\section{Discussion and Conclusions}

The scheme proposed in this paper for the interpretation of quantum
indeterminism offers a scope that allows for an objective ontology of the
physical world besides the possibility of being undetermined. Such a scheme is
more realistic and more consistent than the observer-dependent interpretation
that is implied by the von Neumann and the Everett-Wheeler interpretations.
The re-creation scheme is free of the known paradoxes of quantum measurements
like Schr\"{o}dinger's cat and the EPR since it does not consider a subjective
role for measurements or a wave function collapse. The scheme presented
exhibit a natural presence of entanglement of states belonging to the same
system. This is the direct effect of the re-creation. Moreover, this scheme
resolves the statistical nature of quantum mechanics by allowing the
statistical distribution of the possible values that an observable might take
to fall within the natural process of continued re-creation of that observable.

It is important to note that the above scheme will not affect the standard
calculations of quantum mechanics, except that it might motivates new
investigations into regions which are until now have not been  excavated by
mainstream research works. Examples of these are the existence of macroscopic
quantum states and the possibility of understanding the gamma ray burst being
a result of some macroscopic quantum processes taking place under very
specific conditions deep in the universe. However, the proposed scheme here is
by no means complete and is open for further development.\bigskip

\textbf{ Acknowledgment}

This work was supported by a grant from the John Templeton Foundation.

\end{document}